\documentclass[aps,prd,superscriptaddress,showpacs,amsmath,amssymb]{revtex4}
\usepackage{graphicx}
\usepackage[]{latexsym}
\usepackage{bm}

\newcommand{\be}{\begin{equation}}\newcommand{\ee}{\end{equation}}
\newcommand{\bea}{\begin{eqnarray}}\newcommand{\eea}{\end{eqnarray}}
\newcommand{\brr}{\begin{array}}\newcommand{\err}{\end{array}}
\newcommand{\bit}{\begin{itemize}}\newcommand{\eit}{\end{itemize}}
\newcommand{\ben}{\begin{enumerate}}\newcommand{\een}{\end{enumerate}}

\newcommand{\ba}{\begin{array}}
\newcommand{\ea}{\end{array}}

\def\lan{\langle}
\def\lf{\left}

\def\non{\nonumber}\def\ran{\rangle}

\def\ri{\right}
\def\al{\alpha}

\def\te{\theta}
\def\si{\sigma}
\def\om{\omega}

\def\1{{_{1}}}\def\2{{_{2}}}

\def\noHe0{:\;\!\!\;\!\!:H_e(0):\;\!\!\;\!\!:}
\def\noHm0{:\;\!\!\;\!\!:H_\mu(0):\;\!\!\;\!\!:}

\def\lan{\langle}
\def\lf{\left}

\def\non{\nonumber}
\def\ran{\rangle}

\def\ri{\right}

\def\al{\alpha}

\def\te{\theta}

\def\si{\sigma}
\def\om{\omega}

\def\1{{_{1}}}\def\2{{_{2}}}

\begin{document}

\title{Mixing-induced Spontaneous Supersymmetry Breaking}

\author{Antonio Capolupo}
\affiliation{DMI,
Universit\`a degli Studi di Salerno, Via Ponte don Melillo,
I-84084 Fisciano (SA), Italy} \affiliation{INFN Sezione di Napoli,
Gruppo collegato di Salerno, Italy}

\author{Marco Di Mauro}
\affiliation{DMI,
Universit\`a degli Studi di Salerno, Via Ponte don Melillo,
I-84084 Fisciano (SA), Italy} \affiliation{INFN Sezione di Napoli,
Gruppo collegato di Salerno, Italy} \affiliation{FNSPE, Czech Technical University in Prague,
B\v{r}ehov\'a 7, 115 19 Praha 1, Czech Republic}

\author{Alfredo Iorio}
\affiliation{Faculty of Mathematics and Physics, Charles University in Prague,
V Hole\v{s}ovi\v{c}k\'ach 2, 18000 Praha 8, Czech Republic}



\begin{abstract}
It is conjectured that flavor mixing furnishes a universal mechanism for the spontaneous breaking of supersymmetry. The conjecture is proved explicitly for the mixing of two Wess--Zumino $\mathcal{N}=1$ supermultiplets and arguments for its general validity are given. The mechanism relies on the fact that, despite mixing treats fermions and bosons symmetrically, both the fermionic and the bosonic zero point energies are shifted by a positive amount and this kind of shift does not respect supersymmetry.
\end{abstract}

\maketitle

Supersymmetry (SUSY) is a posited symmetry of nature that has many nice features. Its breaking has been the subject of an intense investigation in particle physics in the last decades because, since SUSY is not observed, it must be either broken or ruled out as a fundamental symmetry of nature. The first models exhibiting spontaneous SUSY breaking were proposed in \cite{Fayet:1974jb,O'Raifeartaigh:1975pr}, while the dynamical breaking was discussed in \cite{Witten:1981nf}.

An (apparently) separated and important aspect of modern particle physics is flavor mixing, which occurs in both the hadronic \cite{Cabibbo:1963yz, Kobayashi:1973fv} and leptonic \cite{Bilenky:1978nj} sectors of the Standard Model.

In this letter we conjecture that flavor mixing, even when it occurs between supermultiplets (hence preserving SUSY at the Lagrangian level) always induces a spontaneous breaking \footnote{We do not consider here the trivial cases of a mixing that explicitly breaks SUSY.}. Since mixing is seen experimentally, our conjecture points at flavor mixing as the responsible for the fact that SUSY is not seen at our scales.

The conjecture is based on the following argument. First of all we recall that, given any supersymmetric theory, the SUSY algebra implies that a sufficient condition for spontaneous SUSY breaking is a nonzero vacuum energy (see, e.g., \cite{Witten:1981nf}). In Majorana notation this is written as follows (in this letter we adopt the conventions of \cite{Wess:1974tw,Wess:1973kz}):
\bea
\langle 0|H|0\rangle = \frac{1}{8} \langle 0| \textrm{Tr}(C\gamma^0\lf[Q,Q\ri]_{+})|0\rangle \neq  0 \qquad \Rightarrow \qquad Q_{\alpha}|0\rangle \neq 0.
\eea
where $H$ is the Hamiltonian of the theory, $Q$ is the associated supercharge and $C$ is the charge conjugation matrix.

Now consider for definiteness the Lagrangian $\mathcal{L}(\psi_1, B_1;\psi_2, B_2)$ for two massive supermultiplets $\Phi_i$ with fermionic components $\psi_i$ and bosonic components $B_i$, with $i=1,2$ and with $m_1\neq m_2$. We have:
\bea\label{SUSYsign}
\langle 0|H|0\rangle=0,
\eea
which implies $Q_{\alpha}|0\rangle = 0$. Here $|0\rangle=|0\rangle_1\otimes|0\rangle_2$ is the vacuum state for $H$, the Hamiltonian corresponding to $\mathcal{L}$. Eq.(\ref{SUSYsign}) is a necessary signature of SUSY, and it is due to the fact that (here and in the following we set $\hbar=c=1$):
\bea\label{Canc}
H=H_{\psi}+H_B\sim\sum_{\mathbf{k},i} \,\lf\{  \omega_{k,i}^\psi\lf( N_{k,i}^\psi - \frac{1}{2}  \ri)+ \omega_{k,i}^B \lf( N_{k,i}^B + \frac{1}{2}  \ri) \ri\},
\eea
and, being $\omega_{k,i}^\psi=\omega_{k,i}^B \equiv\omega_{k,i}$, we have $\omega_{k,i} \lf( \frac{1}{2}-\frac{1}{2}\ri) =0$, i.e. the zero point energies cancel. For the sake of clarity in (\ref{Canc}) we consider the free fields case but, of course, this cancelation is model independent.

Let us suppose that the two supermultiplets are mixed as follows:
\bea\label{Mixing}
\Phi_a &=& \cos\theta\,\, \Phi_1 + \sin\theta\,\,\Phi_2 =G^{-1}(\theta)\Phi_1 G(\theta); \\ \non
\Phi_b &=& -\sin\theta\,\, \Phi_1 + \cos\theta\,\,\Phi_2 =G^{-1}(\theta)\Phi_2 G(\theta).
\eea
In this way mixing does not break SUSY at the Lagrangian level. On the far right hand sides we have written the transformations in terms of a generator $G(\theta)$.
When the fields are mixed there is a new vacuum state, called the \emph{flavor vacuum} \cite{Blasone:1995zc,Alfinito:1995kx,Blasone:2001du,Ji:2002tx}, given by:
\bea
|0\rangle_f=G^{-1}(\theta)|0\rangle
\eea
Our conjecture states that the expectation value of the Hamiltonian on this vacuum is always of the form:
\bea\label{SUSYbreak}
_f\langle 0|H|0\rangle_f = h(\theta,m_1,m_2)\geq 0
\eea
with $h=0$ only if $\theta=0$ or $m_1=m_2$, hence SUSY is broken by the nontrivial mixing. Notice that the vacuum energies here are those of particles of definite mass (mass eigenstates) as computed on the flavor vacuum, that is why there is no problem in principle to write the corresponding $\omega_{k,i}$s (for details on this point see, e.g., \cite{Blasone:1995zc,Alfinito:1995kx,Blasone:2001du,Ji:2002tx} and later in this letter for the explicit case of the massive Wess-Zumino model).

The effect shown by (\ref{SUSYbreak}) relies on the fact that, despite mixing treats fermions and bosons symmetrically in (\ref{Mixing}), both the fermionic and the bosonic zero point energies are shifted by a positive amount. This kind of shift does not respect SUSY which would require a negative shift for the fermionic zero point energy combined with a positive shift of the same magnitude for the bosonic zero point energy, or viceversa (see Fig.~1 below). Let us now prove it explicitly for a simple case.

Consider the Wess--Zumino Lagrangian for two massive free supermultiplets whose field content is two free Majorana fermions $\psi_i$, two free real scalars $S_i$, two free real pseudoscalars $P_i$, two scalar dummy fields $F_i$ and two pseudoscalar dummy fields $G_i$, ($i=1,2$):
\bea\label{lagrangianoff}
\mathcal{L} = - \frac{i}{2}\,\bar{\psi}_1\,\not\!\partial \psi_1  - \frac{1}{2}\,\partial_{\mu}S_1 \partial^{\mu}S_1 \,
- \frac{1}{2}\,\partial_{\mu}P_1 \,\partial^{\mu}P_1 \,
+\,\frac{1}{2}\, F_1^2 \,+\, \frac{1}{2}\,G_1^2
+ m_1 \lf(F_1 S_1 + G_1 P_1 - \frac{i}{2} \bar{\psi}_1 \,\psi_1\ri) + (1\rightarrow 2) \;.
\eea
The field content of the theory is the same as that of a $\mathcal{N}=2$ hypermultiplet, which is made out of two $\mathcal{N}=1$ Wess--Zumino supermultiplets. Nevertheless, the $SU(2)_R$ symmetry for us is explicitly broken by the request of having $m_1\neq m_2$, hence we are not dealing with a $\mathcal{N}=2$ SUSY theory but rather with two copies of a $\mathcal{N}=1$ Wess--Zumino theory. For this reason we consider the following SUSY transformations that leave invariant (up to a surface term, as customary) the off-shell Lagrangian (\ref{lagrangianoff}):
\bea
\delta S_i &=& i{\bar \alpha}\psi_i \;, \;\; \delta P_i = i{\bar \alpha} \gamma_5 \psi_i; \label{SUSYtransf1}\\
\delta \psi_i &=& \partial_{\mu}(S_i - \gamma_5 P_i) \gamma^{\mu}\alpha + (F_i + \gamma_5 G_i)\alpha; \label{SUSYtransf2}\\
\delta F_i &=& i{\bar \alpha} \gamma^\mu \partial_{\mu} \psi_i \;, \; \; \delta G_i = i{\bar \alpha} \gamma_5 \gamma^\mu \partial_{\mu} \psi_i \label{SUSYtransf3}
\eea
where $i=1,2$, and the parameter $\alpha$ is a Majorana spinor.

We keep the off-shell formulation for the sake of showing that SUSY is untouched by the mixing at the Lagrangian level (see later). Of course, the dynamical information resides in the on-shell expressions that are recovered by using the Euler-Lagrange equations for the dummy fields,
\bea\label{dummyonmass}
F_i^{on}=-m_i S_i \;, \; \; G_i^{on}=-m_i P_i, \qquad i=1,2 \;,
\eea
in the Lagrangian (\ref{lagrangianoff}) to obtain
\bea\non\label{lagrangian}
\mathcal{L} &=&- \frac{i}{2}\,\bar{\psi}_1\,(\not\!\partial + m_1)\,\psi_1 \,\, -\, \frac{1}{2}\,\partial_{\mu}S_1 \, \partial^{\mu}S_1 \,-\,
\frac{1}{2}\, m_1^2 \,S_1^2
\,- \frac{1}{2}\,\partial_{\mu}P_1 \,\partial^{\mu}P_1 \,-\, \frac{1}{2}\,m_1^2 \,P_1^2 + (1\rightarrow 2)
\\
 &=& - \frac{i}{2}\,\bar{\psi}\,( \not\!\partial + M_d)\,\psi \,-\,
 \frac{1}{2}\,\partial_{\mu}S \,\partial^{\mu}S \,-\,
 \frac{1}{2} \,S^{T}\, M_d^2 \,S \,-\,
  \frac{1}{2}\,\partial_{\mu}P \,\partial^{\mu}P \,-\,\frac{1}{2}\, P^{T} \,M_d^2 \,P\,,
 \eea
(here $\psi=(\psi_1,\psi_2)^T, S=(S_1,S_2)^T, P=(P_1,P_2)^T$, and $M_d=\textrm{diag}(m_1,m_2)$) and in the expressions for the transformations (\ref{SUSYtransf2}) and (\ref{SUSYtransf3}). Note that the transformations (\ref{SUSYtransf3}) become identities satisfied when the fermions are on-shell, while (\ref{SUSYtransf2}) becomes
\bea
\delta \psi_i = \partial_{\mu}(S_i - \gamma_5 P_i) \gamma^{\mu}\alpha  - m_i (S_i + \gamma_5 P_i)\alpha \label{SUSYtransf4} \;.
\eea
The transformations (\ref{SUSYtransf1}) are untouched by this procedure. The on-shell Lagrangian (\ref{lagrangian}) is invariant under (\ref{SUSYtransf1}) and (\ref{SUSYtransf4}).

By applying the usual Noether method to the off-shell Lagrangian (\ref{lagrangianoff}) we find the following conserved SUSY current associated to the transformations (\ref{SUSYtransf1}) - (\ref{SUSYtransf3}):
\bea \label{susycurrent}
J^{\mu}=\sum_{i=1}^{2} [ \gamma^{\lambda}\partial_{\lambda}(S_i - \gamma_5 P_i)\gamma^{\mu}\psi_i + m_i\gamma^{\mu}(S_i - \gamma_5 P_i) \psi_i ] \;.
\eea
Note that the dummy fields go on-shell automatically in the procedure and the same conserved current is obtained by using the on-shell expressions (\ref{lagrangian}) and (\ref{SUSYtransf1}) and (\ref{SUSYtransf4}). The latter is a universal feature of SUSY-Noether currents \cite{Iorio:1999yx,Iorio:2000xv} and we can make use of it anytime.

Consider now the mixing transformations (\ref{Mixing}) for the model (\ref{lagrangianoff}):
\bea\label{mixing-transf}
\psi_f= U \psi, \qquad S_f = U S,\qquad P_f= U P, \qquad F_f = U F,\qquad G_f= U G,
\eea
where $\psi_f=(\psi_a, \psi_b)^T$ , etc., $F = (F_1,F_2)^T$, $G = (G_1,G_2)^T$, and $U= \lf(\begin{array}{cc} \cos\theta&\sin\theta\\ -\sin\theta &\cos \theta\end{array}\ri)$. With these the Lagrangian (\ref{lagrangianoff}) becomes
\bea
\mathcal{L}_f & = &- \frac{i}{2}\,\bar{\psi}_a\,\not\!\partial \psi_a  - \frac{1}{2}\,\partial_{\mu}S_a \partial^{\mu}S_a \,
- \frac{1}{2}\,\partial_{\mu}P_a \,\partial^{\mu}P_a \,
+\,\frac{1}{2}\, F_a^2 \,+\, \frac{1}{2}\,G_a^2
+ m_a \lf(F_a S_a + G_a P_a - \frac{i}{2} \bar{\psi}_a \,\psi_a\ri) + (a\rightarrow b) \non \\
&+& m_{a b} \lf(F_a S_b + F_b S_a + G_a P_b +  G_b P_a - \frac{i}{2} \bar{\psi}_a \,\psi_b - \frac{i}{2} \bar{\psi}_b \,\psi_a\ri) \label{Mixlagrangianoff} \;,
\eea
where $m_a = m_1\cos^2\theta + m_2\sin^2\theta$, $m_b = m_1\sin^2\theta + m_2\cos^2\theta$, and $m_{ab}=(m_2-m_1)\sin\theta\cos\theta$, while,
due to linearity of both the SUSY transformations (\ref{SUSYtransf1})-(\ref{SUSYtransf3}) and the mixing transformations (\ref{mixing-transf}) we have ($\sigma=a,b)$:
\bea
\delta S_\sigma &=& i{\bar \alpha}\psi_\sigma \;, \;\; \delta P_\sigma = i{\bar \alpha} \gamma_5 \psi_\sigma; \label{MixSUSYtransf1}\\
\delta \psi_\sigma &=& \partial_{\mu}(S_\sigma - \gamma_5 P_\sigma) \gamma^{\mu}\alpha + (F_\sigma + \gamma_5 G_\sigma)\alpha; \label{MixSUSYtransf2}\\
\delta F_\sigma &=& i{\bar \alpha} \gamma^\mu \partial_{\mu} \psi_\sigma \;, \; \; \delta G_\sigma = i{\bar \alpha} \gamma_5 \gamma^\mu \partial_{\mu} \psi_\sigma \label{MixSUSYtransf3}
\eea
i.e. off-shell the mixed fields transform just like the unmixed ones under SUSY. The first terms of ${\cal L}_f$ in (\ref{Mixlagrangianoff}) (first line) are invariant in form with respect to the unmixed Lagrangian (\ref{lagrangianoff}) or, in other words: ${\cal L}_f = {\cal L} + {\cal L}_{mix}$, where all functions are evaluated in the new variables $S_f$, etc.. This is as it must be, because the mixing transformations are not a symmetry of $\mathcal{L}$ but rather a {\it redefinition} of the fields, which in general changes the functional form of the Lagrangian $\mathcal{L} \to \mathcal{L}_f$. Another well known example of such a redefinition is the shift of the fields to the ``true vacuum'' in theories with spontaneous symmetry breaking.

On the other hand, it might appear that the form invariance of the SUSY transformations (\ref{MixSUSYtransf1})-(\ref{MixSUSYtransf3}) only guarantees the invariance of the first line in (\ref{Mixlagrangianoff}) descending directly from the invariance of (\ref{lagrangianoff}) under (\ref{SUSYtransf1})-(\ref{SUSYtransf3}). In other words, it appears that ${\cal L}_{mix}$ explicitly breaks SUSY. In fact, it is not so. It is straightforward to see that ${\cal L}_{mix}$ is invariant on its own under (\ref{MixSUSYtransf1}) - (\ref{MixSUSYtransf3}) , $\delta {\cal L}_{mix} = \partial_\mu (i \bar{\alpha} m_{a b} [(S_b + \gamma_5 P_b) \gamma^\mu \psi_a + (a \leftrightarrow b)])$, hence the conserved SUSY current is
\bea \label{mixsusycurrent}
J_f^{\mu}=\sum_{\sigma=a}^{b} [ \gamma^{\lambda}\partial_{\lambda}(S_\sigma - \gamma_5 P_\sigma)\gamma^{\mu}\psi_\sigma
+ m_\sigma \gamma^{\mu}(S_\sigma - \gamma_5 P_\sigma) \psi_\sigma
+ m_{\sigma \tau} \gamma^{\mu}(S_\tau - \gamma_5 P_\tau) \psi_\sigma] \;,
\eea
with $\tau = a,b$ and $\tau \neq \sigma$. As for the Lagrangian, we can write $J^\mu_f = J^\mu + J^\mu_{mix}$, with obvious notation but keeping in mind that all functions are evaluated in the new variables $S_f$, etc., i.e. it is crucial to notice, that, just like $\mathcal{L} (S, ...) = {\mathcal L}_f (S_f, ...)$ also for the currents $J^\mu (S, ...) = J^\mu_f (S_f, ...)$. It is easy to convince oneself that the difference between the functional form of $J^\mu$ in (\ref{susycurrent}) and of $J^\mu_f$ above resides in the different expressions for the mixed dummy fields on-shell \cite{Iorio:1999yx, Iorio:2000xv} obtained from (\ref{Mixlagrangianoff}) (compare with (\ref{dummyonmass}))
\bea\label{dummyonflavor}
F_\sigma^{on}=-m_\sigma S_\sigma - m_{\sigma \tau} S_\tau \;, \; \; G_\sigma^{on}=-m_\sigma P_\sigma - m_{\sigma \tau} P_\tau  \;,
\eea
where $\sigma, \tau = a,b$ and $\tau \neq \sigma$ as the on-shell expression for the fermionic transformation (\ref{MixSUSYtransf2}) is now (compare with (\ref{SUSYtransf4}))
\bea
\delta \psi_\sigma = \partial_{\mu}(S_\sigma - \gamma_5 P_\sigma) \gamma^{\mu}\alpha
- m_\sigma (S_\sigma + \gamma_5 P_\sigma)\alpha
- m_{\sigma \tau} (S_\tau + \gamma_5 P_\tau)\alpha \label{MixSUSYtransf4} \;,
\eea
with $\sigma, \tau = a,b$ and $\tau \neq \sigma$ (the other transformations (\ref{MixSUSYtransf1}) and (\ref{MixSUSYtransf3}) have exactly the same fate as the corresponding ones in the unmixed case (\ref{SUSYtransf1}) and (\ref{SUSYtransf3}), respectively). From this we conclude that
the on-shell Lagrangian obtained by using (\ref{dummyonflavor}) in (\ref{Mixlagrangianoff}) has the following form
\bea\label{MixLagrangian}
\mathcal{L}_f= -\frac{i}{2}\,\bar{\psi}_f\,( \not\!\partial + M)\,\psi_f \,-\, \frac{1}{2}\,
\partial_{\mu}S_f \,\partial^{\mu}S_f \,
-\,\frac{1}{2} \,S_f^{T}\, M^2 \,S_f \,-\, \frac{1}{2}\,\partial_{\mu}P_f \,\partial^{\mu}P_f \,-\,
\frac{1}{2} \,P_f^{T} \,M^2 \,P_f,
\eea
with $M=\lf(\begin{array}{cc} m_a&m_{ab}\\ m_{ab} &m_b\end{array}\ri)$, is left invariant by the on-shell transformations (\ref{MixSUSYtransf1}) and (\ref{MixSUSYtransf4}) and the conserved current is $J^\mu_f$ in (\ref{mixsusycurrent}). Notice that (\ref{MixLagrangian}) coincides with the Lagrangian obtained by implementing the mixing transformations directly on the on-shell unmixed Lagrangian (\ref{lagrangian}), as it should be.

As seen a crucial role is played here by the linearity of SUSY transformations. That is the reason why the mixing transformations (\ref{mixing-transf}) commute with the SUSY transformations (\ref{SUSYtransf1})-(\ref{SUSYtransf3}) allowing for SUSY to be preserved in the mixed Lagrangian (\ref{MixLagrangian}). This is a very robust result and can be easily applied to more general situations. For instance, interaction terms can be accommodated with no changes in this formalism without spoiling such a linearity. Indeed, we just need to work with dummy fields both in the Lagrangian and in the transformation rules, as we did here for the massive free Wess-Zumino case, and only later move to the dynamical fields.

Having established the SUSY of the mixed Lagrangian, in what follows we shall not need the off-shell expressions and we shall refer only to the on-shell Lagrangians, fields and transformations. We shall compute the expectation value of the Hamiltonian on the flavor vacuum to prove that SUSY is spontaneously broken. To that end we need the Fourier expansion of the fields ($i=1,2$):
\bea
\psi_i(x)&=& \sum_{r=1}^2\int \frac {d^3\mathbf{k}}{(2\pi)^{\frac{3}{2}}}\,\, e^{i \mathbf{k}\mathbf{x}}\lf[u^r_{\mathbf{k},i}(t)\alpha^r_{\mathbf{k},i}
+ v^r_{-\mathbf{k},i}(t)\alpha^{\dagger
r}_{\mathbf{-k},i}\ri],\\
S_i(x)&=& \int \frac {d^3\mathbf{k}}{(2\pi)^{\frac{3}{2}}}\,\, \frac{1}{\sqrt{2 \omega_{k,i}}}  e^{i \mathbf{k}\mathbf{x}}\lf[b_{\mathbf{k},i} e^{-i\omega_{k,i}t}
+ b^{\dagger}_{\mathbf{-k},i}e^{i\omega_{k,i}t}\ri],\\
P_i(x)&=& \int \frac {d^3\mathbf{k}}{(2\pi)^{\frac{3}{2}}}\,\, \frac{1}{\sqrt{2 \omega_{k,i}}}  e^{i \mathbf{k}\mathbf{x}}\lf[c_{\mathbf{k},i} e^{-i\omega_{k,i}t}
+ c^{\dagger}_{\mathbf{-k},i}e^{i\omega_{k,i}t}\ri],
\eea
where $v^r_{\mathbf{k},i}=\gamma_0 C (u^r_{\mathbf{k},i})^*$ and $u^r_{\mathbf{k},i}=\gamma_0 C (v^r_{\mathbf{k},i})^*$ by the Majorana condition  and the operators $\alpha^r_{\mathbf{k},i}$, $b_{\mathbf{k},i}$ and $c_{\mathbf{k},i}$ annihilate the vacuum $|0\rangle=|0\rangle^{\psi}\otimes|0\rangle^S\otimes|0\rangle^P$. Clearly the expectation value of the Hamiltonian on this vacuum is zero, as proved by a straightforward computation:
\bea
\langle 0|H_{\psi}|0\rangle = -
\,  \int d^{3}{\bf k} \, (\omega_{k,1} + \omega_{k,2}),
\eea
and
\bea
\langle 0|H_B|0\rangle =
\,  \int d^{3}{\bf k} \, (\omega_{k,1} + \omega_{k,2}),
\eea
where $H_B=H_S+H_P$, so that:
\bea
\langle 0|(H_{\psi}+ H_B)|0\rangle=0.
\eea
The mixing transformations
(\ref{mixing-transf}) can be written as:
\bea
\label{mixing-Generator}
\psi_{\si}(x)\equiv G^{-1}_{\psi }(\te) \; \psi_{i}(x)\; G_{\psi }(\te)\,,
\qquad
S_{\si}(x)\equiv G^{-1}_S(\te) \; S_{i}(x)\; G_S(\te)\,,
\qquad
P_{\si}(x)\equiv G^{-1}_{P}(\te) \; P_{i}(x)\; G_{P}(\te)\,,
\eea
respectively, where $(\si,i)=(a,1), (b,2)$,
and $G^{-1}_{\psi }(\te)$, $G^{-1}_S(\te)$, $G^{-1}_{P}(\te)$,
are the generators of the mixing transformations \cite{Blasone:1995zc}-\cite{Blasone:2001du},\cite{Capolupo:2004}, which are given by:
\bea
G_{\psi}(\theta) &=& \exp\lf[\frac{\theta}{2}\int d^3 \mathbf{x}\lf(\psi_1^{\dagger}(x)\psi_2(x) - \psi_2^{\dagger}(x)\psi_1(x)\ri) \ri];\\
G_S(\theta) &=& \exp\lf[-i\theta \int d^3 \mathbf{x}(\pi_1^S(x) S_2(x)- \pi_2^S(x) S_1(x)) \ri];\\
G_P(\theta) &=& \exp\lf[-i\theta \int d^3 \mathbf{x}(\pi_1^P(x) P_2(x)- \pi_2^P(x) P_1(x)) \ri],
\eea
where with $\pi_i^S(x)$ and $\pi_i^P(x)$ we have denoted the conjugate momenta of the fields $S_i(x)$ and $P_i(x)$, respectively.

The flavor
annihilation operators are defined as $\al^{r}_{{\bf k},\si}
\equiv G^{-1}_{\psi}(\te)\;\al^{r}_{{\bf k},i} \;G_{\psi}(\te)$,
   $ b_{{ \bf k},\si}\equiv
 G^{-1}_S(\te)\; b_{{ \bf k},i}\;
G_S(\te),$ and $ c_{{ \bf k},\si} \equiv
 G^{-1}_{P}(\te)\; c_{{ \bf k},i}\;
G_{P}(\te).$  They annihilate the flavor vacuum
$|0\ran_{f}\,\equiv\,|0\ran_{f}^{\psi}\, \otimes \,|0\ran_{f}^S\,
\otimes \,|0\ran_{f}^{P}\, $, where:
\bea
\label{mixed-vacua}
|0\ran_{f}^{\psi}\,\equiv \, G^{-1}_{\psi }(\te) \; |0\ran^{\psi}\,,
\qquad
|0\ran_{f}^S\,\equiv \, G^{-1}_S(\te) \; |0\ran^S\,,
\qquad
|0\ran_{f}^{P}\,\equiv \, G^{-1}_{P}(\te) \; |0\ran^{P}\,,
\eea
are the flavor vacua of the fields $\psi_{\si}(x)$, $S_{\si}(x)$, $P_{\si}(x)$,
respectively.

The crucial point  of our discussion is that the vacuum $|0\ran_{f}$ is a (coherent) condensate:
\bea\label{condensate1}
 {}_{f}\langle 0| \al_{{\bf k},i}^{r \dag} \al^r_{{\bf
k},i} |0\ran_{f} & = & \sin^{2}\te ~ |V^{\psi}_{{\bf
k}}|^{2},
\\\label{condensate2}
{}_{f}\langle 0| b_{{\bf k},i}^{ \dag} b_{{\bf
k},i} |0\ran_{f} & = &   {}_{f}\langle 0| c_{{\bf k},i}^{\dag}
 c_{{\bf k},i} |0\ran_{f} \,=\, \sin^{2}\te ~ |V^B_{{\bf
k}}|^{2},
\eea
where $i=1,2$ and the reference frame in which ${\bf k}=(0,0,|{\bf k}|)$ has
been adopted for convenience.  The moduli of the two functions $V^{\psi}_{{\bf k}}$ and  $V^B_{{\bf k}}$
are given by:
\bea\label{Bogoliubov}
|V^{\psi}_{{\bf k}}|  =  \frac{ (\om_{k,1}+m_{1}) - (\om_{k,2}+m_{2})}{2
\sqrt{\om_{k,1}\om_{k,2}(\om_{k,1}+m_{1})(\om_{k,2}+m_{2})}}\, |{\bf k}| \, ,
\qquad
| V^B_{{\bf k}}|  =   \frac{1}{2} \lf( \sqrt{\frac{\om_{k,1}}{\om_{k,2}}} -
\sqrt{\frac{\om_{k,2}}{\om_{k,1}}} \ri)\,.
\eea

The expectation value of the fermionic part of $H$ is given by \cite{Blasone:2004yh}:
\bea\label{Hflav}{}_{f}\lan
0| H_{\psi} | 0 \ran_{f}\, =-\,  \int d^{3}{\bf k} \,
(\omega_{k,1} + \omega_{k,2}) \,(1 - 2\,|V^{\psi}_{\bf k}|^{2}
\sin^{2}\theta)\, ,  \eea
while for the bosonic part we obtain:
\bea\label{HflavBos} {}_{f}\lan
0| H_B | 0 \ran_{f}\, =\,  \int d^{3}{\bf k} \,
(\omega_{k,1} + \omega_{k,2}) \,(1 + 2\,|V^B_{\bf k}|^{2}
\sin^{2}\theta)\,. \eea
Combining Eqs.(\ref{Hflav}) and (\ref{HflavBos}) we have the result:
\bea\label{violation}
\, {}_{f}\lan 0| (H_{\psi}\,+\, H_B) | 0 \ran_{f}\,=\,
2\,\sin^{2}\theta \, \int d^{3}{\bf k} \,
(\omega_{k,1} + \omega_{k,2}) \,(|V^{\psi}_{\bf k}|^{2} + \,|V^B_{\bf k}|^{2})
\,,
\eea
which is different from zero and positive when $\theta\neq0$ and $m_1\neq m_2$. This proves our conjecture for the case in point. As announced, the reason of this spontaneous SUSY breaking is that the mixing shifts both the fermionic and the bosonic zero point energies by a positive amount, hence it does not respect SUSY as shown schematically in Fig.~1.

 \begin{figure}
 \centering
  \includegraphics[height=.2\textheight]{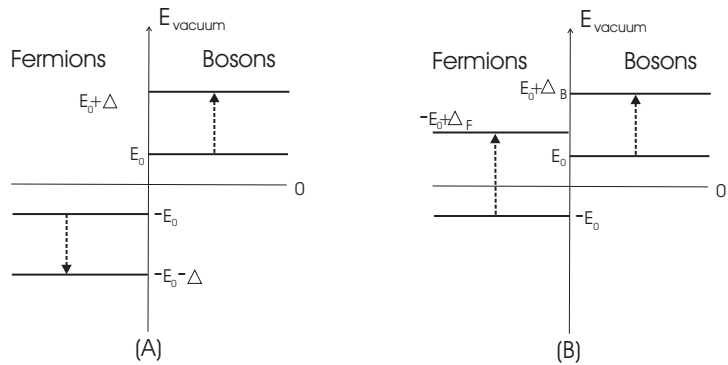}
  \caption{(A) SUSY preserving shifts of the zero-point energies. (B) Schematic behaviors of the mixing-induced shifts that break SUSY in the vacuum. Notice that the breaking is present also in the case that $\Delta_F = \Delta_B$.}
\label{shift}
\end{figure}

Of course, this is only a very partial proof of the conjecture but it is reasonable to argue that the mechanism shown here for the Wess--Zumino model is actually model-independent at least when SUSY can be implemented linearly hence the mixing transformations commute with it. Indeed, the qualitative features of the mixing of higher spin fields are the same as those presented here \cite{Ji:2002tx}: in the flavor vacuum those fields condense and shift the zero point energy always of a positive amount, whether they are bosons or fermions. Having shown here that the breaking mechanism induced by mixing is based precisely on this ``SUSY asymmetric'' vacuum condensate effect of Fig.~1 and it does not depend on the explicit form of the functions in (\ref{Bogoliubov}), since in (\ref{Hflav})-(\ref{violation}) only their squares appear, we can attempt to say that supermultiplets of higher spin should not alter the final outcome. Another possible issue with the general case is whether interacting fields condense in the flavor vacuum in a way that spoils the spontaneous breaking. Unfortunately, the explicit expressions of the functions corresponding to those in (\ref{Bogoliubov}) in this case are not easy to obtain but it seems very unlikely that interaction could restore SUSY in the vacuum by modifying the condensates from the free case of Fig.~1 (B) to that of Fig.~1 (A).

Having clarified that, a more urgent task than a complete proof of the conjecture (which, nonetheless, is surely one direction worth investigating) perhaps would be to probe our conjecture within phenomenologically relevant models. This latter program is the way to test whether it is realistic to consider mixing as the actual responsible for SUSY breaking.

Other directions to investigate are the connection of the SUSY breaking illustrated here with the well known SUSY breaking induced by a nonzero temperature \cite{Das:1997gg} (in a different context a connection between the mixing angle and temperature has been proposed in \cite{Blasone:2010zn}); and the possible cosmological implications of Eq.(\ref{violation}) (in a non supersymmetric context it has been shown that the flavor vacuum energy can be interpreted as a new dark energy component of the universe \cite{Capolupo:2006et}, \cite{Capolupo:2008rz}). Further analysis of these aspects will be done elsewhere.

\section*{Acknowledgements}

A.~C. and M.~D. thank M.~Blasone and P.~Jizba for useful discussions.

\end{document}